\documentclass[prd,onecolumn,preprintnumbers,nofootinbib]{revtex4}
\usepackage{natbib}
\usepackage[plainpages=false, colorlinks=true, anchorcolor=blue, linkcolor=blue, citecolor=blue, bookmarks=false]{hyperref}

\usepackage{amsmath,upgreek}
\usepackage{graphicx}%
\usepackage{multirow}%
\usepackage{amsmath,amssymb,amsfonts}%

\begin{document}

\title{A meta-analysis of impact factors of particle physics journals using NASA/ADS}
%\titlerunning{Stability of impact factors of Astronomical journals}

\author{Rayani Venkat Sai Rithvik} \altaffiliation{E-mail: ee21btech11043@iith.ac.in}
\author{Shantanu Desai}
\altaffiliation{E-mail: shntn05@gmail.com}

\affiliation{$^{1}$Department of Electrical Engineering,
IIT Hyderabad,   Telangana-500084 India}

\affiliation{$^{2}$Department of Physics, IIT Hyderabad,   Telangana-500084 India}

\begin{abstract}
In this work, we calculate the 2025 impact factors for  16 journals in Particle Physics (and related areas such as Nuclear Physics)
using citations collated by NASA/ADS (Astrophysics Data System).
We then compare them to the official
impact factors calculated by Clarivate. We also compare these impact factors to the median-based impact factors introduced in a previous work. We do not find any systematic  bias between the ADS and official impact factors. We find the maximum relative difference between official and median-based impact factors for PTEP and Annual Review of Nuclear and Particle Science. For PTEP, this difference is due to one outlier, viz Particle Data Group with over 1400 citations in the last two years. Similarly for  Annual Review of Nuclear and Particle Science, the difference is due to a review paper on Hubble tension and early dark energy with over 100 citations in the past two years. Both these papers have significantly elevated the  official impact factors above its median value.

 \end{abstract}

\maketitle

%%\fundingInfo{Funding info text.}

%\maketitle

%\footnotetext{\footnote{\url{https://en.wikipedia.org/wiki/Impact_factor}}}

\section{Introduction}
In a previous work~\cite{Rithvik24} (R25 hereafter), we  calculated   the impact factors of about 40 Astrophysics journals  using NASA/ADS~\cite{ADS}. We also studied the stability of these impact factors in a follow-up work~\cite{Rithvik2}. We had  also introduced  in R25,  a new median-based impact factor and compared it with the traditional impact factors  for all the journals. We briefly recap the definitions of both these impact factors.

The  traditional way to calculate the impact factor, (which we referred to as Old Impact Factor in R25) in year \( n \) is defined as the ratio of the total number of citations in year \( n-1 \) of all papers published in the journal during years \( n-2 \) and \( n-3 \), divided by the number of refereed papers published in those same years~\footnote{\url{https://en.wikipedia.org/wiki/Impact_factor}}:

\begin{equation}
\text{Impact Factor}_{\text{old}}(n) = \frac{C_{n-1}}{P_{n-2} + P_{n-3}}
\label{eq:IF}
\end{equation}
where:
\begin{itemize}
    \item \( C_{n-1} \) is the total number of  refereed+ unrefereed citations in year \( n-1 \) to papers published in years \( n-2 \) and \( n-3 \).
    \item \( P_{n-2} \) is the number of refereed publications in the journal in year \( n-2 \).
    \item \( P_{n-3} \) is the number of refereed publications in the journal in year \( n-3 \).
\end{itemize}
Therefore, the impact factor of a journal in the year  2025 is equal to the 
total citations received in 2024 for all papers published in that journal in  2023 and 2022.
 We note that the official impact factors which are calculated by Clarivate~\footnote{https://mjl.clarivate.com/home},   are based on the citations obtained using Web of Science bibliometric data.
%These citations have sometimes been refered to called the Science Citation Index (SCI)~\citep{Frogel}
%The impact factor has been used to judge the relative importance of journals and could be used by  hiring  and proposal funding committees.

%It has been pointed out that sometimes citations are missed or not attributed to the right person~\citep{Will}. Furthermore, errors in citations collated by Institute for Scientific Information  have been noted, because of non-standard conventions used by astronomers~\citep{Abt04}. Therefore, we re-evaluate the impact factors using citations collated by NASA/ADS~\citep{ADS}, which is the definitive resource and database for all astrophysics publications and compare them to the official impact factors.

The second impact factor (introduced in R25)  based on the  median number of citations is refereed  to as  \textbf{New Impact Factor}.
The \textbf{New Impact Factor} in year \( n \) is defined as the median number of citations in year \( n-1 \) of all refereed papers published in the journal during years \( n-2 \) and \( n-3 \), divided by the total number published papers during years \( n-2 \) and \( n-3 \). %We then  used NASA/ADS~\citep{ADS} to calculate the new and old impact factor for about 38 of the most widely used journals in astrophysics.

In this work, we now carry out the same exercise as R25 for Particle Physics journals, which publish papers in the broad areas of High Energy, Nuclear, and Particle Physics. The manuscript is structured as follows. Our analysis and results are described in Sect.~\ref{sec:results} and we conclude in Sect.~\ref{sec:conclusions}.

\section{ Analysis and Results}
\label{sec:results} 
Similar to R25, we obtained the citations using the NASA/ADS API available at \url{https://ui.adsabs.harvard.edu/help/api/} and checked the particle physics publications by using the tag {\tt collections:Physics}. We note that although the inspirehep database~\cite{inspirehep} is the standard reference used by the high energy physics community, we are not aware of a corresponding API available from inspire-hep to obtain information on citations and related metadata with sufficient documentation. Furthermore, NASA/ADS also collates papers (and corresponding citations) outside of astronomy in all journals where particle physics papers are published. Therefore, similar to R25 we use NASA/ADS based citations for our analysis. We considered 16 journals (cf. Table~\ref{tab:tab1}) which accept papers in Particle Physics and related areas such as Nuclear Physics. Some of these journals were also previously analyzed for astrophysics papers in R25.
Some of the analyzed journals such as PRD, PTEP, EPJC, Internal Journal of Modern Physics A also accept papers in astrophysics, general relativity,  and  cosmology, whereas some of them such as Europhysics Letters, Modern Physics Letters A, EPJP, PhRvL consider papers in all areas of Physics.  We also note that many papers in the area of Cosmology and Gravitational Waves, which are primarily classified as astrophysics papers in arXiv, are also cross-listed under the ``Physics'' tag in NASA/ADS and would get included in our analysis.
Finally, we note that NASA/ADS collection of citations is not always complete
and is known to contain false positives~\cite{Will}. 

With the above caveats in mind, we report
our results for the old and new impact factors for 16 journals. A tabular summary can be found in Table~\ref{tab:tab1}.  The first column contains the journal abbreviations used in ADS. The full names of the journals can be found in the Appendix. The second column contains the official impact factor obtained using Clarivate. The third and fourth columns contain results for the old and new impact factors computed using NASA/ADS.  In addition to these impact factors, we also collated miscellaneous publication and citations related statistics, similar to that done in R25. These include the total number of published papers in 2022 and 2023, fraction of papers without citations, and the number of citations for the most cited paper for each journal. 
These can be found in Table~\ref{tab:tab2}. We also note that none of these journals have Article Processing charges, as long as one opts for subscription-based access. Therefore, unlike R25 details of APC are redundant here.
Our observations based on the results are as follows:
\begin{itemize}
\item The official impact factors are in agreement with the ADS-calculated impact factors for almost all journals.  Unlike the astrophysics journals R25, we do not find a systemic increase or any other bias with respect to the official impact factors, since about half of the journals have higher ADS-based impact factors, whereas the remaining have a higher official impact factor. The maximum difference is about 2.5 for ARNPS (Annual Review of Nuclear and Particle Science).
\item  The new and old impact factors agree for most journals. The two exceptions are ARNPS and PTEP with an increase in the old impact factor of 5.5 (for both), corresponding to a fractional increase of 143\% and 
553\% respectively.
\item For PTEP, the top two cited papers, namely Ref.~\cite{PDG} and Ref.~\cite{LiteBird} citations of 1412 and 108, respectively The first citation corresponds to Particle Data Group (PDG). Therefore, PDG citation (almost 200 times larger than the ADS based average impact factor) has significantly elevated the impact factor for PTEP. That is why there is a large difference between the new and old impact factors. We note that the second-top-cited paper in PTEP by the LiteBird Collaboration is actually an astophysics/cosmology paper, but it has been cross-listed in Physics by NASA/ADS. However, some of its science goals  such as to probe the energy scale of inflation are pertinent to Particle Physics. We note that PTEP also showed a significant difference between these two impact factors for Astrophysics papers~\cite{Rithvik24}.
\item Similarly for ARNPS, the paper with the highest citations for 2023 and 2024 is 
Ref.~\cite{Riess} with 107 citations which is almost 25 times larger than the median impact factor.  This paper is a review of Hubble tension and early dark energy and straddles Astrophysics, Particle Physics, and General Relativity~\cite{Riess}.
Therefore, the large difference between the two impact factors for these two journals is due to the above two outliers in terms of citations.
\item The journals having maximum fraction of papers with no citations are PTEP, Modern Physics Letters A and Europhysics Letters, corresponding to about 45\%. 
 \end{itemize}

\begin{table}[t]
\begin{tabular}{|c|c|c|c|}
\hline
\textbf{Journal Code} & \textbf{Official Impact factor} & \textbf{(Old) Impact Factor using ADS} &  \textbf{New Impact Factor using ADS}    \\ 
  \hline
  ARNPS & 8.4  & 10.97 & 4.5 \\ 
  \hline
  EL &1.8  & 1.52 & 1 \\ \hline
  EPJA & 2.8  & 2.98 & 2\\ \hline
  EPJC & 4.99  & 5.07 & 3 \\  \hline
  EPJP & 2.9  & 1.75 & 1 \\ \hline
  IJMPA & 1.2 & 1.30 & 0 \\ \hline
 JHEP & 5.5 & 6.30 & 6 \\ \hline
  JPhG & 3.5  &3.93 &3 \\ 
  \hline
  MPLA  & 1.6 & 1.51 & 1 \\ 
  \hline
  NIMPA  & 1.4  & 0.98 & 0 \\ 
  \hline
  NuPHB  & 2.8  & 2.66 & 1 \\ 
  \hline
  PhLB & 4.5  &  4.48 & 3 \\ 
  \hline
  PhRvC & 3.4 & 3.67 & 2 \\ \hline
  PhRvD & 5.3   & 5.91  & 3 \\ 
  \hline
  PhRvL & 9.0  & 8.91  & 5 \\ \hline
  PTEP & 8.3  & 6.53 & 1 \\ 
  \hline
  \end{tabular}
 \caption{\label{tab:tab1} Summary of old and new impact factors of particle Physics journals The first column indicates the acronym used for this journal in NASA/ADS. The full  names of these journals can be found in Table~\ref{tab2}} 
\end{table}

 \begin{table}[!htbp]
\begin{tabular}{|c|c|c|c|}
\hline
\textbf{Journal Code} &  \textbf{ \# Published (2022+2023)} &  \textbf{Fraction of Papers with no citations} & \textbf{Citations of top cited paper}    \\ 
\hline
 ARNPS & 34  & 0.0882 & 107 \\ 
  \hline
  EL & 748  & 0.4479 & 36 \\ \hline
  EPJA & 570  & 0.2789 & 64\\ \hline
  EPJC & 2345  & 0.1872 & 247 \\  \hline
  EPJP & 2551  & 0.4042 & 31 \\ \hline
  IJMPA & 585 & 0.5162 & 17 \\ \hline
 JHEP & 4997 & 0.1011 & 115 \\ \hline
  JPhG & 243 & 0.2428 & 74 \\ 
  \hline
  MPLA  & 455 & 0.4418 & 18 \\ 
  \hline
  NIMPA  & 1846  & 0.5114 & 25 \\ 
  \hline
  NuPHB  & 605  & 0.2992 & 33 \\ 
  \hline
  PhLB & 1400  &  0.1771 & 52 \\ 
  \hline
  PhRvC & 1889 & 0.1752 & 44 \\ \hline
  PhRvD & 8320   & 0.1131  & 307 \\ 
  \hline
  PhRvL & 4321  & 0.0569  & 343 \\ \hline
  PTEP & 343  & 0.4344 & 1412 \\ \hline
  \end{tabular}
  \caption{\label{tab:tab2} Some miscellaneous publication related statistics for  the 16 journals discussed in Table~\ref{tab:tab1}.} 
  \end{table}

\section{Conclusions}
\label{sec:conclusions}
In this work, we have done an extensive meta-analysis of citations for 16 of the  most widely used Particle   Physics journals
along the same lines as our previous work R25. To the best of our knowledge this is the first such study  of citations of Particle Physics journals.
We  independently calculated the 2025  impact factors of 16 journals which accept Particle Physics and related areas such as Nuclear Physics based  papers, using NASA/ADS database. We then compared them to the official impact factor of each journal, which have been obtained using the SCI based citations calculated by Clarivate. We then compared these with a median-based impact factor introduced in R25. Our results for the same can be found in Table~\ref{tab:tab1}, while some ancillary publications/citation-related statistics can be found in Table~\ref{tab:tab2}.

Unlike astrophysics journals, we did not find any systematic bias in the ADS based impact factors compared to the official impact factors. We also find that the new impact factors are comparable to the old impact factors for all journals except PTEP and Annual Review of Nuclear and Particle Science, which have a net increase of 143\% and 555\%, respectively, in the impact factors.  For PTEP, this is due to the PDG related paper which has 1412 citations~\cite{PDG}. For Annual Review of Nuclear and Particle Science, the citations of a review paper on Hubble tension and early dark energy~\cite{Riess} with 107 citations.

\iffalse
\section*{Data availability}
There are no data associated with the article. The codes for analysis can be made available upon a reasonable request to the corresponding author.
\fi

\bibliography{main}%

%\nocite{*}% Show all bib entries - both cited and uncited; comment this line to view only cited bib entries;

%\bibliographystyle{apalike}
\section*{Appendix}

\begin{table}[h]
\begin{tabular}{|c|c|}
\hline
 \textbf{Journal Code} & \textbf{Journal Name}   \\ 
  \hline
  ARNPS & Nuclear and Particle Science  \\ \hline
   EL & Europhysics Letters  \\
  \hline
  EPJA &  The European Physical Journal A \\ \hline
  EPJC & The European Physical Journal C \\ \hline
  EPJP & European Physical Journal Plus \\ \hline
  IJMPA & International Journal of Modern Physics A \\ \hline
  JHEP & Journal of High Energy Physics \\ \hline
  JPHG & Journal of Physics G \\ \hline
  MPLA & Mod. Physics Letters A.\\ \hline
  NIMPA & Nuclear Instruments and Methods A \\ \hline
  NuPHB  & Nuclear Physics B \\ \hline
  PHLB & Physics Letters B \\ 
  \hline
  PhRvC & Physical Review C  \\  \hline
  PhRvD & Physical Review D  \\  \hline
  PhRvL & Physical Review Letters \\
  \hline 
  PTEP  & Progress of Theoretical and Experimental Physics  \\ 
  \hline
\end{tabular}
\caption{Names of journals  corresponding to the NASA/ADS codes analyzed in Table~\ref{tab:tab1} and Table~\ref{tab:tab2}.
\label{tab2}}
\end{table}
\end{document}